\documentclass[conference]{IEEEtran}
\IEEEoverridecommandlockouts
\usepackage{cite}
\usepackage{amsmath,amssymb,amsfonts}
\usepackage{algorithmic}
\usepackage{graphicx}
\usepackage{textcomp}
\usepackage{xcolor}
\def\BibTeX{{\rm B\kern-.05em{\sc i\kern-.025em b}\kern-.08em
    T\kern-.1667em\lower.7ex\hbox{E}\kern-.125emX}}
\begin{document}

\title{Hybrid Reconfigurable Intelligent Surface Enabled Over-the-Air Index-Modulation\\
}

\makeatletter
\newcommand{\linebreakand}{%
  \end{@IEEEauthorhalign}
  \hfill\mbox{}\par
  \mbox{}\hfill\begin{@IEEEauthorhalign}
}
\makeatother

\author{

\author{\IEEEauthorblockN{Emre Arslan\IEEEauthorrefmark{1}\IEEEauthorrefmark{3}, Ali Tugberk Dogukan\IEEEauthorrefmark{2}\IEEEauthorrefmark{3}, Fatih Kilinc\IEEEauthorrefmark{2}\IEEEauthorrefmark{3}, Ertugrul Basar\IEEEauthorrefmark{3}}

\IEEEauthorblockA{\IEEEauthorrefmark{1}6GEN Laboratory, Next-Generation R\&D, Network Technologies, Turkcell, {\.{I}}stanbul, Turkiye}

\IEEEauthorblockA{\IEEEauthorrefmark{2}Department of Research and Development, ULAK Communications Inc., {\.{I}}stanbul, Turkiye}

\IEEEauthorblockA{\IEEEauthorrefmark{3}Department of Electrical and Electronics Engineering, Koç University, {\.{I}}stanbul, Turkiye\\ Emails: {emre.arslan@turkcell.com.tr},{\{fatih.kilinc,tugberk.dogukan\}@ulakhaberlesme.com.tr},{ebasar@ku.edu.tr}}}

\IEEEauthorblockN{1\textsuperscript{st} Emre Arslan}
\IEEEauthorblockA{\textit{Next Generation Research and Development }\\
\textit{6GEN. Lab, Turkcell Iletisim Hizmetleri Inc.}\\
Istanbul, Turkey \\ 
emre.arslan@turkcell.com.tr}
\and
\IEEEauthorblockN{2\textsuperscript{nd} Ali Tugberk Dogukan}
\IEEEauthorblockA{\textit{Department of Research and Development} \\
\textit{ULAK Communications Inc.}\\
Istanbul, Turkey \\
tugberk.dogukan@ulakhaberlesme.com.tr}

\linebreakand
\IEEEauthorblockN{3\textsuperscript{rd} Fatih Kilinc}
\IEEEauthorblockA{\textit{ Department of Research and Development} \\
\textit{ ULAK Communications Inc.}\\
Istanbul, Turkey \\
fatih.kilinc@ulakhaberlesme.com.tr}
\and
\IEEEauthorblockN{4\textsuperscript{th} Ertugrul Basar}
\IEEEauthorblockA{\textit{Department of Electrical and Electronics} \\
\textit{Engineering, Ko\c{c} University}\\
Istanbul, Turkey \\
ebasar@ku.edu.tr} 
}

\maketitle

\begin{abstract}
Users' desire for enhanced performance drive the inevitable technological progress on vertical applications in wireless communication systems. To meet these demands, researchers vigorously investigate potential 6G and beyond technologies and solutions. Reconfigurable intelligent surfaces (RISs) have risen in popularity and attracted the attention of academia as well as industry and seem to be a promising candidate for 6G technology. RISs are reflective metamaterials with many configurable elements consisting of pin diodes that have the ability to manipulate the impinging signals' properties, hence, enabling some sort or virtual control over the wireless channel. In this paper, we propose a novel over-the-air index modulation (IM) scheme through the use of a hybrid RIS with active and passive partitions to convey additional IM bits over-the-air. In addition we propose another modified scheme that allows the transmission for even more IM bits but at the expense of certain trade-offs. First, we present the system model for both of the proposed schemes. Furthermore, comprehensive computer simulation results are provided and discussed presenting the superior bit error rate (BER) performance and additional benefits of the proposed systems compared to similar benchmarks.

\end{abstract}


\section{Introduction}

\begin{figure*}[t]
		\centering
		\includegraphics[width=1.0 \textwidth]{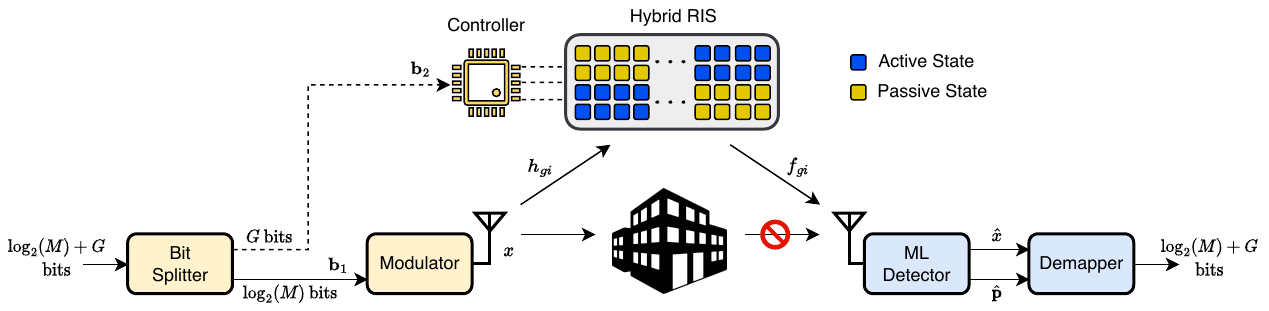} 
		\caption{ \centering System model of OTA-RIS-IM.}
		\label{fig:Fig1}
\end{figure*}  

\IEEEPARstart{W}{ireless} communication technologies are rapidly developing as researchers migrate from one generation (5G) to another (6G), striving to meet users' stringent demands as a consequence of imminent vertical applications. Numerous candidate solutions and novel technologies are proposed in the literature each manifesting high potential for future wireless communication systems. While some methods are adopted, most of these are dismissed and are not favored due to distinct drawbacks and critical issues which may be in terms of practicality, performance, complexity and more. However, as new hot topics arise, researchers must revert to previous solutions and consider them and their integration with recently developed and considered technologies which may provide new insights and together provide superior performance. Hence, previously dismissed technologies may be rekindled as once again significantly promising solutions. 

Index modulation (IM) is a simple, spectrum and energy efficient scheme to embed and transmit additional information bits through the indices of resource blocks in communication systems \cite{IM,basar2017index,wen2017index,tugtekin2023coordinate}. These resource blocks on which information is embedded can be subcarriers, antennas, time-slots, constellation, activation orders and  much more \cite{sugiura2017state,wen2017index}. It is a very popular concept with a significant amount of research and studies conducted on it and has proven itself to be a worthy tool and is considered for future communication networks. IM shows benefits such as the ability to improve error performance, provide enhanced energy and spectral efficiency, and all while providing low complexity \cite{cheng2018index}. One of the most important characteristic of IM is its flexibility and compatibility attribute since it can be employed and coexist in a wireless communication system in numerous methods \cite{sugiura2017state, mao2018novel}. One example is orthogonal frequency division multiplexing-IM (OFDM-IM) where IM is incorporated in the subcarriers of an OFDM waveform as an addition layer of modulation by activating certain subcarriers to transmit additional bits over OFDM subblock \cite{bacsar2013orthogonal, wen2015achievable}.
In \cite{arslan2020index} and \cite{csahin2021ofdm}, IM has been integrated with non-orthogonal multiple access and joint radar and communication technologies introducing novel system designs for future technologies; proposing a flexible OFDM-IM subcarrier activation and power allocation method according to users' needs and a unique waveform to serve both communication and radar functionalities, respectively. Furthermore, additional interesting studies have been conducted such as \cite{dualmodeIM} where a novel technique is proposed by dividing the subcarriers into two groups and modulating them with distinguishable mode constellations and \cite{arslan2020sparse} where bits are transmitted through the indexing of multiple codebooks to be used in sparse vector coding for ultra-reliable  low-latency communication networks.

Reconfigurable intelligent surfaces (RISs) have recently risen in popularity in the wireless communication community and exhibit high potential as a promising candidate for future 6G and beyond wireless communication systems. RISs are large reflective surfaces with numerous small and low-cost meta-material elements that have the ability to manipulate the characteristics of impinging signals. Hence, RISs are able to reflect, refract, absorb and even amplify impinging signals without the need for additional buffering or processing, providing some sort of virtual control over the wireless channel. Generally, RISs are envisioned to be employed for coverage enhancement, increased signal quality, elevated capacity and other system performance improvements \cite{subrt2012controlling, huang2020holographic, kilinc2021physical, su2020capacity, perovic2020channel, yigit2022hybrid}. One significant advantage of the RIS technology is its flexibility in the integration with conventional technologies. There are many studies that present substantial benefits in the integration of RIS with previously considered technologies such as non-orthogonal multiple accessing \cite{arslan2022reconfigurable}, machine learning \cite{huang2019indoor}, localization \cite{ma2020indoor}, OFDM \cite{pradhan2020reconfigurable}, mutiple-input and multiple-output \cite{tang2020mimo}, and much more \cite{9614985, arslan2023network, arslan2022over}.

Considering both the advantages and flexibility of RIS and IM, the integration of these technologies are being investigated to further improve wireless communication system performances. In \cite{dash2022performance}, the authors present a greedy detector for an RIS space-shift keying system where the RIS phases are aligned to maximize the instantaneous signal-to-noise ratio (SNR) at the target receive antenna which is selected through IM bits. Furthermore, an intelligent RIS grouping-based IM (RGB-IM) is proposed to enhance the spectral efficiency and bit error rate (BER) performance of conventional RIS systems \cite{asmoro2022ris}. In this study the RIS is divided in to multiple groups each with the same number of elements, and according to the incoming IM bits a group of RIS elements are activated while the other RIS elements are inactive and introduce zero phase. However, completely inactivating remaining groups and elements are not optimal and could be used in a more efficient manner to improve the reliability and spectral efficiency of the system.

In this study, we propose a novel over-the-air RIS-IM (OTA-RIS-IM) technique exploiting the entire RIS enhancing the spectral efficiency and BER performance. We present a hybrid RIS that has multiple groups and each group can be active and passive according to the IM bits. Furthermore, we further expand this system design and propose an enhanced OTA-RIS-IM (E-OTA-RIS-IM) system by introducing an absorption mode at the RIS to send additional IM bits at the expense of slightly reduced BER performance improvement compared to our benchmark. 

The rest of the paper can be summarized as follows. In Section II, the system framework of the proposed OTA-RIS-IM and E-OTA-RIS-IM scheme is presented. In addition, Section III consists of the computer simulation results and discussions while Section IV concludes the paper.

\begin{figure*}[t]
		\centering
		\includegraphics[width=1.0 \textwidth]{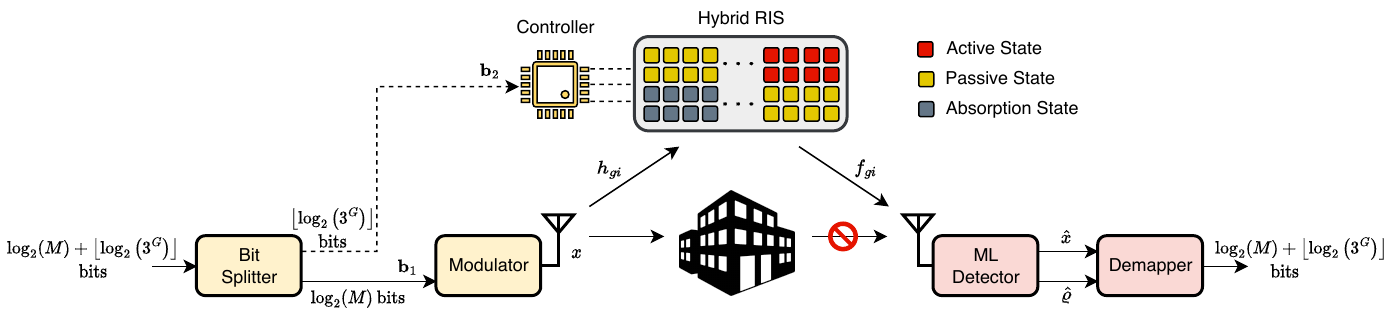} 
		\caption{ \centering System model of E-OTA-RIS-IM.}
		\label{fig:Fig2}
\end{figure*}  

\section{System Model}

In this section, we provide the system model for both of the proposed schemes. First we will present the initial OTA-RIS-IM system design. Furthermore, we will build on top of it to propose the E-OTA-RIS-IM system. As seen from Fig. \ref{fig:Fig1}, RIS includes $N$ elements and is partitioned into $G$ groups, each including $\bar{N}$ elements, $N=G \bar{N}$. Here, the reflection coefficient of the $i$th element of the $g$th RIS group is $\theta_{gi} = e^{j \vartheta_{gi}}$ with unit amplitude, where $\vartheta_{gi} \in \left[0,2\pi\right)$ is the phase shift adjusted by the RIS controller, $i=1,2,\cdots,\bar{N}$, and $g=1,2,\cdots,G$. 

Assume that $L_h = \zeta_0 d_h^{-\varkappa_h}$ and $L_f = \zeta_0 d_f^{-\varkappa_f}$ represent the large scale path loss of the BS-to-RIS and RIS-to-UE, respectively, where $\zeta_0$, is the path loss at the reference distance of 1 meter, and $d_h$,  $d_f$
and $\varkappa_h$, $\varkappa_f$ denote the distance from BS to RIS and RIS to UE, the path loss exponents at BS to RIS and RIS to UE paths, respectively. Then, we define the channel coefficients from the BS to the $\left(g,i \right)$th RIS element and from the $\left(g,i \right)$th RIS element to the UE as $h_{gi}= \left|h_{gi}\right| e^{j \phi_{gi}}=\sqrt{L_h}\bar{h}_{gi}$  and $f_{gi}= \left|f_{gi}\right| e^{j \varphi_{gi}}=\sqrt{L_f}\bar{f}_{gi}$, respectively. 
Moreover, $\Bar{h}_{gi}$ and $\Bar{f}_{gi}$ are independent Rician fading channel coefficients as follows:
\begin{align}
    \Bar{h}_{gi} &= \sqrt{\frac{K_h}{K_h+1}} \Bar{h}_{gi}^{\mathrm{LoS}} + \sqrt{\frac{1}{K_h+1}} \Bar{h}_{gi}^{\mathrm{NLoS}}, \\
    \Bar{f}_{gi} &= \sqrt{\frac{K_f}{K_f+1}} \Bar{f}_{gi}^{\mathrm{LoS}} + \sqrt{\frac{1}{K_f+1}} \Bar{f}_{gi}^{\mathrm{NLoS}},
\end{align}
where $\Bar{h}_{gi}^{\mathrm{LoS}}$ and $\Bar{f}_{gi}^{\mathrm{LoS}}$ represent the line-of-sight (LoS) components and $\Bar{h}_{gi}^{\mathrm{NLoS}}$ and $\Bar{f}_{gi}^{\mathrm{NLoS}}$ represent the non-LoS (NLoS) components of $\bar{h}_{gi}$ and $\bar{f}_{gi}$, respectively, whereas $K_h$ and $K_f$ denote the Rician $K$ factor of the paths between the BS-RIS and the RIS-UE, respectively. Here, $\Bar{h}_{gi}^{\mathrm{LoS}}$, $\Bar{f}_{gi}^{\mathrm{LoS}}$, $\Bar{h}_{gi}^{\mathrm{NLoS}}$, and $\Bar{f}_{gi}^{\mathrm{NLoS}}$ are independent and identically distributed complex Gaussian random variables with $\mathcal{CN}(0,1)$.


\subsection{OTA-RIS-IM}

The system model of OTA-RIS-IM is demonstrated in Fig. \ref{fig:Fig1}. In OTA-RIS-IM,  information bits are transmitted both through the conventional $M$-ary PSK/QAM signal constellation and also the indices of the active/passive states of RIS groups. Firstly, $\mathbf{b}_1 = \left[b_{11}, b_{12}, \cdots, b_{1 \kappa}\right]^{\mathrm{T}}$ bit vector is mapped into the modulated symbol $x$, $x\in\mathcal{M}$, where $\mathcal{M}$ is the $M$-ary PSK/QAM signal constellation and $\kappa = \log_2(M)$. Then, $\mathbf{b}_2 = \left[b_{21}, b_{22}, \cdots, b_{2G}\right]^{\mathrm{T}}$ bit vector determines the indices of active/passive states of RIS groups $\mathbf{p} = [p_1,p_2, \cdots, p_G]^{\mathrm{T}}$, $p_g\in \{0,1\}$, $g=1,2,\dots,G$. Here,  the $g$th element of $\mathbf{p}$ is equal to the $g$th element of $\mathbf{b}_2$, i.e., $p_g = b_{2g}$. If $p_g$ is equal to $0$ ($\Bar{p}_g=1$), then the $g$th RIS group is in passive state, and if $p_g$ is equal to $1$ ($\Bar{p}_g=0$), then the $g$th RIS group is in active state, where $\Bar{p}_g$ represents the logical NOT of $p_g$. Here, the number of ones and zeros in $\mathbf{p}$ is equal to the number of active ($G_A$) and passive ($G_P$) RIS groups, respectively, $G_A + G_P = G$. In Table \ref{table:table1}, an example of bit mapping is given for $G=2$. The spectral efficiency of OTA-RIS-IM in terms of bit per channel use  (bpcu) can be given as:
\begin{equation}
    \eta = \log_2 \left(M \right) + G.
\end{equation}

In OTA-RIS-IM, the $(g, i)$th RIS element ($\theta_{gi}$) is selected in such a way that the phase terms in $h_{gi}$ and $f_{gi}$ are eliminated, i.e., $\vartheta_{gi}=-(\phi_{gi} + \varphi_{gi})$. Hence, the received signal at UE can be given as: 
\begin{align} 
&y = \sqrt{P_t} x \left(  \sum_{g=1}^{G} \left[ \mathcal{H}_g \left(\sqrt{\alpha} p_g  + \Bar{p}_g  \right) + {p}_g w^a_g \right] \right)+  w, 
\\ &\mathcal{H}_g = \sum_{i=1}^{\Bar{N}} h_{gi} \theta_{gi} f_{gi} = \sum_{i=1}^{\Bar{N}} \left|h_{gi}\right| \left|f_{gi}\right|, \nonumber
\\ &w^a_g = \sqrt{\alpha }\sum_{i=1}^{\Bar{N}} \theta_{gi} f_{gi} v_{gi}, \nonumber
\end{align}
where $P_t$, $\alpha$, $w$, and $v_{gi}$ are the transmit power, amplification factor at the active RIS groups, additive white Gaussian noise (AWGN) sample with distribution $\mathcal{CN}(0,N_0)$, and amplifier noise term at the $(g,i)$th RIS element with distribution $\mathcal{CN}(0,V_0)$, respectively. Finally, at the UE, the maximum-likelihood (ML) detection rule can be exploited to decode the received signal as follows:
\begin{equation}
    \label{ML}
    \left(\hat{x}, \hat{\mathbf{p}}\right) = \underset{x \in \mathcal{M}, \mathbf{p} \in \mathcal{P}}{\arg \min} \left\| y - \sqrt{P_t}x \left[ \sum_{g=1}^{G} \mathcal{H}_g \left(\sqrt{\alpha} p_g  + \Bar{p}_g  \right) \right] \right\|^2,
\end{equation}
where $\hat{x}$ and $\hat{\mathbf{p}}$ are the estimates of $x$ and $\mathbf{p}$, respectively, and $\mathcal{P}$ is a set including all possible $\mathbf{p}$ vectors. Here, $M 2^G$ number of iterations are performed in the search space to decode the received signal.

The signal-to-noise ratio (SNR) of the received signal can be given as: 
\begin{equation}
    \gamma = \frac{P_t \left| \sum_{g=1}^{G}  \mathcal{H}_g \left(\sqrt{\alpha} p_g  + \Bar{p}_g  \right) \right|^2}{\alpha V_0 \sum_{g=1}^{G} p_g \sum_{i=1}^{\Bar{N}} \left|\theta_{gi} f_{gi} \right|^2 + N_0}.
\end{equation}

\renewcommand{\arraystretch}{1.25}
\begin{table}[t]
\caption{OTA-RIS-IM Example of Bit Mapping For $G=2$}
\centering
\label{table:table1}
\scalebox{0.9}{
\begin{tabular}{|c|c|c|c|c|c|c|}
\hline
$[b_{21}, b_{22}]$ & $G_A$ & $G_P$ & $[p_{1}, p_{2}]$   & $[\Bar{p}_{1}, \Bar{p}_{2}]$          & \textbf{\begin{tabular}[c]{@{}c@{}}First\\ RIS Group \\ State \end{tabular}} & \textbf{\begin{tabular}[c]{@{}c@{}}Second\\ RIS Group \\ State\end{tabular}}    \\ \hline
$[0, 0]$ & $0$ & $2$ & $[0, 0]$ & $[1, 1]$ & Passive & Passive \\ \hline
$[0, 1]$ & $1$ & $1$ & $[0, 1]$ & $[1, 0]$ & Passive & Active \\ \hline
$[1, 0]$ & $1$ & $1$ & $[1, 0]$ & $[0, 1]$ & Active & Passive \\ \hline
$[1, 1]$ & $2$ & $0$ & $[1, 1]$ & $[1, 1]$ & Active & Active \\ \hline
\end{tabular}}
\end{table}

\subsection{E-OTA-RIS-IM}

In this subsection, we present an enhanced OTA-RIS-IM scheme (E-OTA-RIS-IM) to further improve the number of bits transmitted in the system by introducing an absorption state at the hybrid RIS. The system model of E-OTA-RIS-IM is illustrated in Fig. \ref{fig:Fig2}. Here, according to the incoming bits from the backhaul link, the hybrid RIS will determine whether the partitions will behave in the passive, active or absorption state. As in OTA-RIS-IM, firstly, $\mathbf{b}_1 = \left[b_{11},b_{12},\cdots,b_{1\kappa}\right]^{\mathrm{T}}$ bit vector is used to  obtain the $M$-ary modulated symbol $x$, $\kappa=\log_2(M)$. Let us define a codebook $\mathcal{S}$ which includes all permutations with repetition of active/passive/absorption states of RIS groups, $\mathcal{S}=\left\{ \mathbf{s}_1, \mathbf{s}_2, \cdots, \mathbf{s}_{2^{\varpi}} \right\}$, $\varpi=\left\lfloor \log_2 \left(3^G\right) \right\rfloor$, where $\mathbf{s}_{\varrho}=\left[s_{\varrho 1},s_{\varrho 2}, \cdots, s_{\varrho G}\right]^{\mathrm{T}}$ is the $\varrho$th codeword of set $\mathcal{S}$ and $s_{\varrho \mu} \in \{ \sqrt{\alpha}, 1, 0 \}$, $\varrho=1,2,\cdots,2^{\varpi}$, $\mu=1,2,\cdots,G$. Here, $s_{\varrho \mu}$ being $\alpha$, $1$, and $0$ represent active, passive, and absorption states, respectively. Hence, $\mathbf{b}_2=\left[b_{21},b_{22},\cdots,b_{2\varpi}\right]^{\mathrm{T}}$ bit vector is converted into its decimal value ($\xi$), and $\xi$th codeword ($\mathbf{s}_{\xi}$) of $\mathcal{S}$ is selected. As seen in Table \ref{table:E_OTA_RIS_IM}, an example of states of RIS groups selection is provided for $G=2$. Here, a total number of $3^G$ permutations can be generated; however, only $2^{\varpi}$ can be used to transmit information. As can be seen in the table, for the same number of RIS partitions $G$, we can send an additional bit compared to the previous method. Hence, the spectral efficiency of E-OTA-RIS-IM in terms of bpcu can be given as:
\begin{equation}
    \eta = \log_2 \left(M\right) + \left\lfloor \log_2 \left(3^G\right) \right\rfloor.
\end{equation}

As in OTA-RIS-IM, the $(g, i)$th element of RIS is coherently aligned with the corresponding channel coefficients $h_{gi}$ and $f_{gi}$, i.e., $\vartheta_{gi}=-(\phi_{gi} + \varphi_{gi})$. Hence, the received signal at UE can be written as:
\begin{align} 
&y = \sqrt{P_t} x \left(  \sum_{g=1}^{G} \left[ \mathcal{H}_g  s_{\xi g} + \sigma\left(s_{\xi g}\right) w^a_g \right] \right)+  w, 
\end{align}
where $\sigma\left( \beta \right)$ is a function which provides output as $1$ if $\beta$ equals to $\sqrt{\alpha}$, and $0$ otherwise.

In order to decode the received signal, the ML detector can be used as follows:
\begin{equation}
    \label{ML}
    \left(\hat{x}, \hat{\varrho}\right) = \underset{x \in \mathcal{M}, \varrho = 1,\cdots,2^{\varpi}}{\arg \min} \left\| y - \sqrt{P_t}x \left[ \sum_{g=1}^{G} \mathcal{H}_g s_{\varrho g} \right] \right\|^2.
\end{equation}

Moreover, the SNR of the received signal can be obtained as:
\begin{equation}
    \gamma = \frac{P_t \left| \sum_{g=1}^{G}  \mathcal{H}_g s_{\xi g} \right|^2}{\alpha V_0 \sum_{g=1}^{G} \sigma\left(s_{\xi g}\right) \sum_{i=1}^{\Bar{N}} \left|\theta_{gi} f_{gi} \right|^2 + N_0}.
\end{equation}

\renewcommand{\arraystretch}{1.5}
\begin{table}[t]
\label{table:E_OTA_RIS_IM}
\caption{E-OTA-RIS-IM Example of Bit Mapping For $G=2$}
\centering
\scalebox{0.9}{
\begin{tabular}{|c|c|c|c|c|}
\hline
\textbf{\begin{tabular}[c]{@{}c@{}}Index Bits\\ $[b_{21}, b_{22},b_{23}]$\end{tabular}} & $(\xi)$ & \textbf{Codeword} & \textbf{\begin{tabular}[c]{@{}c@{}}First RIS\\ Group State\end{tabular}} & \textbf{\begin{tabular}[c]{@{}c@{}}Second RIS\\ Group State\end{tabular}} \\ \hline
$[0,0,0]$      &           $0$                                                   & $\mathbf{s}_1 = \left[\sqrt{\alpha}, \sqrt{\alpha}\right]^{\mathrm{T}}$    & Active                                                                   & Active                                                                    \\ \hline
$[0,0,1]$      &           $1$                                                  & $\mathbf{s}_2= \left[\sqrt{\alpha}, 1\right]^{\mathrm{T}}$    & Active                                                                   & Passive                                                                   \\ \hline
$[0,1,0]$      &           $2$                                                   & $\mathbf{s}_3 = \left[\sqrt{\alpha}, 0\right]^{\mathrm{T}}$   & Active                                                                   & Absorption                                                                       \\ \hline
$[0,1,1]$      &           $3$                                                  & $\mathbf{s}_4 = \left[1, \sqrt{\alpha}\right]^{\mathrm{T}}$    & Passive                                                                  & Active                                                                    \\ \hline
$[1,0,0]$      &           $4$                                                  & $\mathbf{s}_5 = \left[1, 1\right]^{\mathrm{T}}$    & Passive                                                                  & Passive                                                                   \\ \hline
$[1,0,1]$      &           $5$                                                  & $\mathbf{s}_6 = \left[1, 0\right]^{\mathrm{T}}$   & Passive                                                                  & Absorption                                                                       \\ \hline
$[1,1,0]$      &           $6$                                                  & $\mathbf{s}_7 = \left[0, 1\right]^{\mathrm{T}}$   & Absorption                                                                      & Active                                                                    \\ \hline
$[1,1,1]$      &           $7$                                                  & $\mathbf{s}_8 = \left[0, 1\right]^{\mathrm{T}}$  & Absorption                                                                      & Passive                                                                   \\ \hline
NaN        &           NaN                                                  & NaN               & Absorption                                                                      & Absorption                                                                       \\ \hline
\end{tabular}}
\end{table}

\section{Simulation Results and Comparisons}

\begin{figure}[t]
    \centering
    \includegraphics[width=0.95\columnwidth]{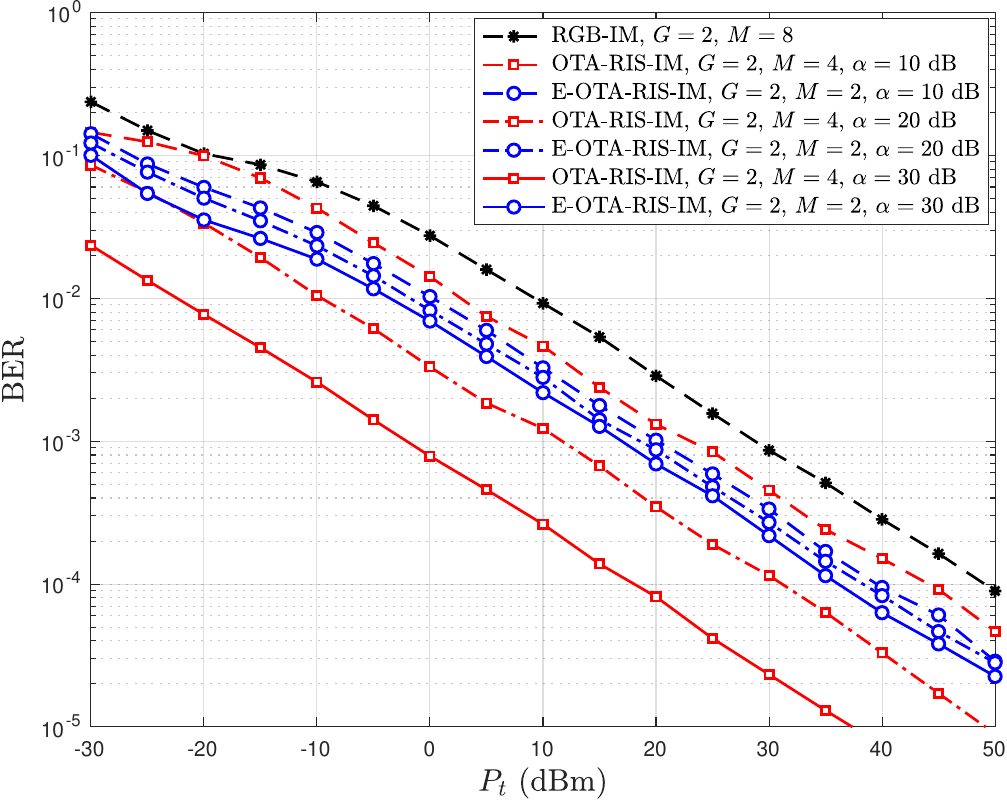}
    \caption{Error performance comparison of OTA-RIS-IM, E-OTA-RIS-IM, and RGB-IM for varying $\alpha$, $N=256$, and $4$ bpcu.}
    \label{fig:number1}
\end{figure}

\begin{figure}[t]
    \centering
    \includegraphics[width=0.95\columnwidth]{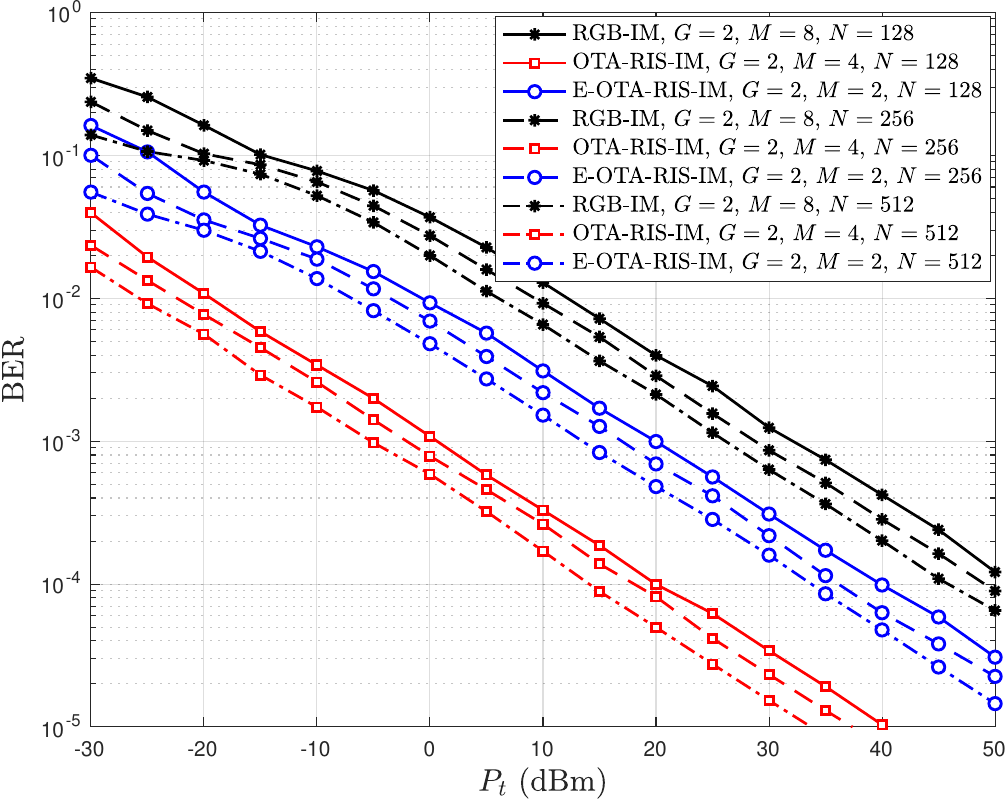}
    \caption{Error performance comparison of OTA-RIS-IM, E-OTA-RIS-IM, and RGB-IM for varying $N$, $\alpha=30$ dB, and $4$ bpcu.}
    \label{fig:number2}
\end{figure}

\begin{figure}[t]
    \centering
    \includegraphics[width=0.95\columnwidth]{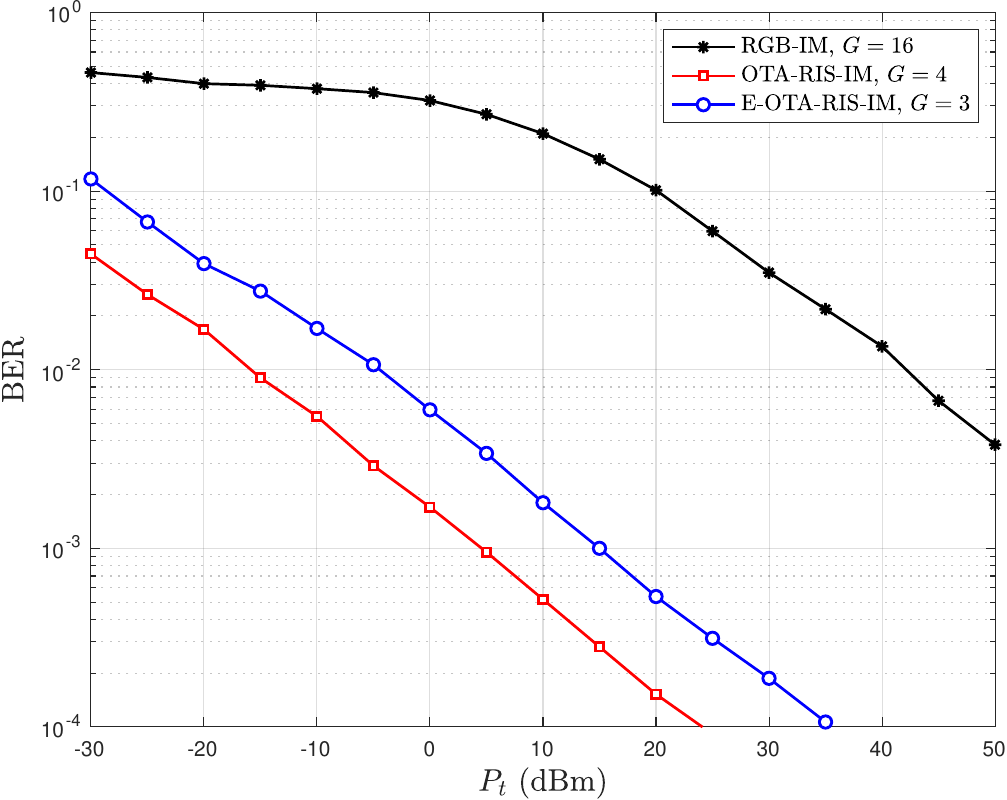}
    \caption{Error performance comparison of OTA-RIS-IM, E-OTA-RIS-IM, and RGB-IM varying $G$, $M=2$, $\alpha=30$ dB, $N=480$, and $5$ bpcu.}
    \label{fig:number3}
\end{figure}

\begin{figure}[t]
    \centering
    \includegraphics[width=0.95\columnwidth]{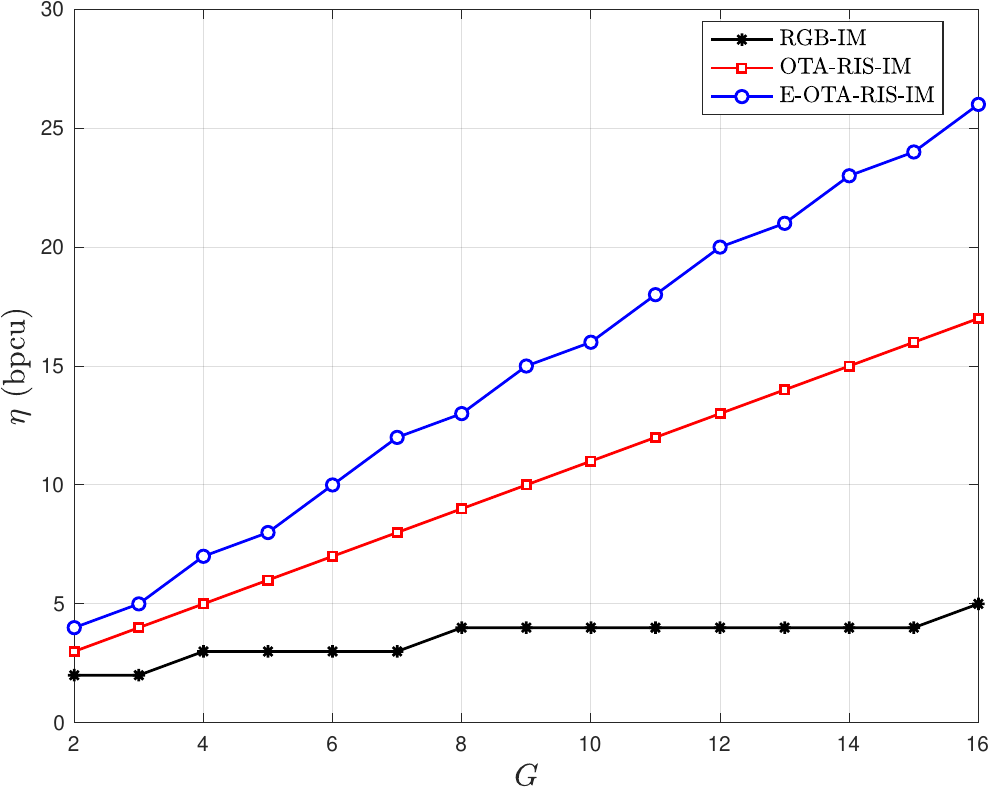}
    \caption{ Spectrall efficiency comparison of all schemes for varying $G$ and $M=2$.}
    \label{fig:number4}
\end{figure}

In this section, we will discuss the simulation results of the proposed OTA-RIS-IM and E-OTA-RIS-IM schemes in terms of BER performance compared to the RGB-IM benchmark \cite{asmoro2022ris} which only activates one group of the RIS. For a different number of RIS groups $G$, various $M$, $N$ and other parameters, we show the superior performance of the proposed scheme over the RGB-IM. Unless otherwise indicated, the following system parameters are assumed in all simulations: the distances $d_h = 20$ and $d_f = 50$ meters, the path loss exponents $\varkappa_h = 2.2$ and $\varkappa_f = 2.8$,
the Rician shape parameters $K_h = K_f = 0$, the noise variances and the reference path loss value of 
$\zeta_0 = -30$ dB. For the BER analyses, the noise power $N_0$ and $V_0$ are fixed to $-130$ dBm and the transmit power $P_t$ is varied. 

Fig. \ref{fig:number1} presents the error performance for all schemes with $G=2$, $N=256$, and $4$ bpcu for varying $\alpha$. It is observed that E-OTA-RIS-IM requires the least modulation order of $M=2$ since it can transmit additional bits through IM and is the most spectral efficient. On the other hand, OTA-RIS-IM needs $M=4$ and RGB-IM benchmark operates at $M=8$ modulation orders to keep the same bpcu of $4$. It is seen that both proposed schemes outperform the benchmark and OTA-RIS-IM provides the best performance of all schemes since it uses the entire RIS at all times due to not having absorption state as in E-OTA-RIS-IM and RGB-IM. Furthermore, the error performance of all schemes improves with respect to $\alpha$.

In Fig. \ref{fig:number2}, we can see the BER performance of the proposed schemes compared to the benchmark with $G=2$ for varying $N$. To obtain same bpcu values in each scheme, the modulation order varies. It is observed that while E-OTA-RIS-IM needs only $M=2$, OTA-RIS-IM needs to operate at $M=4$ modulation order. Despite a higher modulation order, OTA-RIS-IM outperforms E-OTA-RIS-IM in terms of error performance since it has all RIS elements on at all combinations. On the other hand, RGB-IM requires $M=8$ modulation order to keep up the same data rate. It is shown that the RGB-IM at the same bpcu performs significantly worse compared to both of the proposed schemes.

Fig. \ref{fig:number3} presents the BER performance of the proposed systems compared to the benchmark when all of the systems are at the same bpcu and $M=2$. As we can see, to reach the same bpcu RGB-IM requires a large amount of grouping $G$ to send additional bits through IM. On the other hand, OTA-RIS-IM needs only $G=4$ RIS groups to reach the same bpcu and E-OTA-IM needs even a fewer number $G=3$ since it has an additional state the hybrid RIS can take. It can also be observed that since there is no absorption state, even though the bpcu is less compared to E-OTA-RIS-IM, the BER performance is significantly superior because all RIS elements are exploited at all times. 

Fig. \ref{fig:number4} shows the increase in bpcu as the number of groups $G$ increases for each scheme. It can be observed, that as the number of groups increases the benchmark's bpcu does not increase much since only one group is activated according to the $\left\lfloor \log_2(G) \right\rfloor$ IM bits each time and the rest of the RIS is inactive. On the other hand, OTA-RIS-IM uses the entire hybrid-RIS where each group may take an active and passive state according to $G$ IM bits and provides a significantly higher bpcu compared to the benchmark. Furthermore, E-OTA-RIS-IM introduces an absorption state at the RIS to further enhance the bpcu for a trade-off from reliability as we have seen from the previous BER figures. However, we can observe that the bpcu increases drastically for the E-OTA-RIS-IM scheme since the IM bits are increased to $\left\lfloor \log_2(3^G) \right\rfloor$.

\section{Conclusion}

In this study, a novel OTA-RIS-IM and E-OTA-RIS-IM scheme has been proposed, exploiting the benefits of IM and RIS technology efficiently to enhance the spectral efficiency and BER performance of RIS-aided wireless communication systems. Through partitioning and varying the states of the groups at the hybrid-RIS, additional IM bits are conveyed over-the-air. Furthermore, extensive computer simulation results demonstrate and discuss the performance benefits of the proposed system. Future directions may include applications of machine learning, various methods to improve the transmission of IM bits, considering a multiple antenna system design, and other methods to further improve the performance and complexity of the proposed system design.

\section*{Acknowledgment}
This study has been supported by the 1515 Frontier Research and Development Laboratories Support Program of T{\"U}B{\.I}TAK under Project 5229901 - 6GEN. Lab: 6G and Artificial Intelligence Laboratory.

\bibliographystyle{IEEEtran}
\bibliography{References}

\begin{thebibliography}{10}
\providecommand{\url}[1]{#1}
\csname url@samestyle\endcsname
\providecommand{\newblock}{\relax}
\providecommand{\bibinfo}[2]{#2}
\providecommand{\BIBentrySTDinterwordspacing}{\spaceskip=0pt\relax}
\providecommand{\BIBentryALTinterwordstretchfactor}{4}
\providecommand{\BIBentryALTinterwordspacing}{\spaceskip=\fontdimen2\font plus
\BIBentryALTinterwordstretchfactor\fontdimen3\font minus
  \fontdimen4\font\relax}
\providecommand{\BIBforeignlanguage}[2]{{%
\expandafter\ifx\csname l@#1\endcsname\relax
\typeout{** WARNING: IEEEtran.bst: No hyphenation pattern has been}%
\typeout{** loaded for the language `#1'. Using the pattern for}%
\typeout{** the default language instead.}%
\else
\language=\csname l@#1\endcsname
\fi
#2}}
\providecommand{\BIBdecl}{\relax}
\BIBdecl

\bibitem{IM}
E.~{Basar}, ``Index modulation techniques for 5\uppercase{G} wireless
  networks,'' \emph{IEEE Commun. Mag.}, vol.~54, no.~7, pp. 168--175, Jul.
  2016.

\bibitem{basar2017index}
E.~Basar, M.~Wen, R.~Mesleh, M.~Di~Renzo, Y.~Xiao, and H.~Haas, ``Index
  modulation techniques for next-generation wireless networks,'' \emph{IEEE
  Access}, vol.~5, pp. 16\,693--16\,746, Aug. 2017.

\bibitem{wen2017index}
M.~Wen, X.~Cheng, and L.~Yang, \emph{Index modulation for 5\uppercase{G}
  wireless communications}.\hskip 1em plus 0.5em minus 0.4em\relax Springer,
  Jul. 2017, vol.~52.

\bibitem{tugtekin2023coordinate}
O.~F. Tugtekin, A.~T. Dogukan, E.~Arslan, and E.~Basar, ``Coordinate
  interleaved \uppercase{OFDM} with repeated in-phase/quadrature index
  modulation,'' \emph{IEEE Trans. Wirel. Commun.}, May 2023.

\bibitem{sugiura2017state}
S.~Sugiura, T.~Ishihara, and M.~Nakao, ``State-of-the-art design of index
  modulation in the space, time, and frequency domains: Benefits and
  fundamental limitations,'' \emph{IEEE Access}, vol.~5, pp. 21\,774--21\,790,
  Oct. 2017.

\bibitem{cheng2018index}
X.~Cheng, M.~Zhang, M.~Wen, and L.~Yang, ``Index modulation for 5\uppercase{G}:
  Striving to do more with less,'' \emph{IEEE Wirel. Commun.}, vol.~25, no.~2,
  pp. 126--132, Dec. 2018.

\bibitem{mao2018novel}
T.~Mao, Q.~Wang, Z.~Wang, and S.~Chen, ``Novel index modulation techniques: A
  survey,'' \emph{IEEE Commun. Surv. Tutor.}, vol.~21, no.~1, pp. 315--348,
  Jul. 2018.

\bibitem{bacsar2013orthogonal}
E.~Ba{\c{s}}ar, {\"U}.~Ayg{\"o}l{\"u}, E.~Panay{\i}rc{\i}, and H.~V. Poor,
  ``Orthogonal frequency division multiplexing with index modulation,''
  \emph{IEEE Trans. Signal Process.}, vol.~61, no.~22, pp. 5536--5549, Aug.
  2013.

\bibitem{wen2015achievable}
M.~Wen, X.~Cheng, M.~Ma, B.~Jiao, and H.~V. Poor, ``On the achievable rate of
  \uppercase{OFDM} with index modulation,'' \emph{IEEE Trans. Signal Process.},
  vol.~64, no.~8, pp. 1919--1932, Nov. 2015.

\bibitem{arslan2020index}
E.~Arslan, A.~T. Dogukan, and E.~Basar, ``Index modulation-based flexible
  non-orthogonal multiple access,'' \emph{IEEE Wirel. Commun. Lett.}, vol.~9,
  no.~11, pp. 1942--1946, Jul. 2020.

\bibitem{csahin2021ofdm}
M.~M. {\c{S}}ahin, I.~E. Gurol, E.~Arslan, E.~Basar, and H.~Arslan,
  ``\uppercase{OFDM-IM} for joint communication and radar-sensing: a promising
  waveform for dual functionality,'' \emph{Front. Commun. Netw.}, vol.~2, p.
  715944, Aug. 2021.

\bibitem{dualmodeIM}
T.~Mao, Z.~Wang, Q.~Wang, S.~Chen, and L.~Hanzo, ``Dual-mode index modulation
  aided \uppercase{OFDM},'' \emph{IEEE Access}, vol.~5, pp. 50--60, Aug. 2016.

\bibitem{arslan2020sparse}
E.~Arslan, A.~T. Dogukan, and E.~Basar, ``Sparse-encoded codebook index
  modulation,'' \emph{IEEE Trans. Veh. Technol.}, vol.~69, no.~8, pp.
  9126--9130, May 2020.

\bibitem{subrt2012controlling}
L.~Subrt and P.~Pechac, ``Controlling propagation environments using
  intelligent walls,'' in \emph{Proc. Eur. Conf. Antennas Propag.
  ({EUCAP})}.\hskip 1em plus 0.5em minus 0.4em\relax IEEE, May 2012, pp. 1--5.

\bibitem{huang2020holographic}
C.~Huang, S.~Hu, G.~C. Alexandropoulos, A.~Zappone, C.~Yuen, R.~Zhang,
  M.~Di~Renzo, and M.~Debbah, ``Holographic \uppercase{MIMO} surfaces for
  \uppercase{6G} wireless networks: Opportunities, challenges, and trends,''
  \emph{IEEE Wirel. Commun.}, vol.~27, no.~5, pp. 118--125, Jul. 2020.

\bibitem{kilinc2021physical}
F.~Kilinc, I.~Yildirim, and E.~Basar, ``Physical channel modeling for
  \uppercase{RIS}-empowered wireless networks in sub-6 \uppercase{GH}z bands,''
  in \emph{Proc. 55th Asilomar Conf. Signals, Syst., Comput.}, Nov. 2021, pp.
  704--708.

\bibitem{su2020capacity}
R.~Su, L.~Dai, J.~Tan, M.~Hao, and R.~MacKenzie, ``Capacity enhancement for
  irregular reconfigurable intelligent surface-aided wireless communications,''
  in \emph{Proc. IEEE Global Commun. Conf.}\hskip 1em plus 0.5em minus
  0.4em\relax IEEE, Feb. 2020, pp. 1--6.

\bibitem{perovic2020channel}
N.~S. Perovi{\'c}, M.~Di~Renzo, and M.~F. Flanagan, ``Channel capacity
  optimization using reconfigurable intelligent surfaces in indoor mmwave
  environments,'' in \emph{Proc. IEEE Int. Conf. Commun. (ICC)}.\hskip 1em plus
  0.5em minus 0.4em\relax IEEE, Jun. 2020, pp. 1--7.

\bibitem{yigit2022hybrid}
Z.~Yigit, E.~Basar, M.~Wen, and I.~Altunbas, ``Hybrid reflection modulation,''
  \emph{IEEE Trans. Wirel. Commun.}, Nov. 2022.

\bibitem{arslan2022reconfigurable}
E.~Arslan, F.~Kilinc, S.~Arzykulov, A.~T. Dogukan, A.~Celik, E.~Basar, and
  A.~M. Eltawil, ``Reconfigurable intelligent surface enabled over-the-air
  uplink \uppercase{NOMA},'' \emph{IEEE Trans. Green Commun. Netw.}, Dec. 2022.

\bibitem{huang2019indoor}
C.~Huang, G.~C. Alexandropoulos, C.~Yuen, and M.~Debbah, ``Indoor signal
  focusing with deep learning designed reconfigurable intelligent surfaces,''
  in \emph{IEEE Workshop Signal Process. Adv. Wirel. Commun. (SPAWC).}\hskip
  1em plus 0.5em minus 0.4em\relax IEEE, Jul. 2019, pp. 1--5.

\bibitem{ma2020indoor}
T.~Ma, Y.~Xiao, X.~Lei, W.~Xiong, and Y.~Ding, ``Indoor localization with
  reconfigurable intelligent surface,'' \emph{IEEE Commun. Lett.}, vol.~25,
  no.~1, pp. 161--165, Sept. 2020.

\bibitem{pradhan2020reconfigurable}
C.~Pradhan, A.~Li, L.~Song, J.~Li, B.~Vucetic, and Y.~Li, ``Reconfigurable
  intelligent surface \uppercase{(RIS)}-enhanced two-way \uppercase{OFDM}
  communications,'' \emph{IEEE Trans. Veh. Technol.}, vol.~69, no.~12, pp.
  16\,270--16\,275, May 2020.

\bibitem{tang2020mimo}
W.~Tang, J.~Y. Dai, M.~Z. Chen, K.-K. Wong, X.~Li, X.~Zhao, S.~Jin, Q.~Cheng,
  and T.~J. Cui, ``\uppercase{MIMO} transmission through reconfigurable
  intelligent surface: System design, analysis, and implementation,''
  \emph{IEEE J. Sel. Areas Commun.}, vol.~38, no.~11, pp. 2683--2699, Jul.
  2020.

\bibitem{9614985}
M.~Alayasra and H.~Arslan, ``\uppercase{IRS}-enabled beam-space channel,''
  \emph{IEEE Trans. Wirel. Commun.}, vol.~21, no.~6, pp. 3822--3835, Nov. 2022.

\bibitem{arslan2023network}
E.~Arslan, F.~Kilinc, E.~Basar, and H.~Arslan, ``Network-independent and
  user-controlled \uppercase{RIS}: An experimental perspective,'' in
  \emph{Proc. 26th Int. Symp. Wireless Pers. Multimedia Commun. (WPMC)}.\hskip
  1em plus 0.5em minus 0.4em\relax IEEE, Nov. 2023, pp. 1--6.

\bibitem{arslan2022over}
E.~Arslan, I.~Yildirim, F.~Kilinc, and E.~Basar, ``Over-the-air equalization
  with reconfigurable intelligent surfaces,'' \emph{IET Commun.}, vol.~16,
  no.~13, pp. 1486--1497, May 2022.

\bibitem{dash2022performance}
S.~P. Dash, R.~K. Mallik, and N.~Pandey, ``Performance analysis of an index
  modulation-based receive diversity \uppercase{RIS}-assisted wireless
  communication system,'' \emph{IEEE Commun. Lett.}, vol.~26, no.~4, pp.
  768--772, Jan. 2022.

\bibitem{asmoro2022ris}
K.~Asmoro and S.~Y. Shin, ``\uppercase{RIS} grouping based index modulation for
  6\uppercase{G} telecommunications,'' \emph{IEEE Wirel. Commun. Lett.},
  vol.~11, no.~11, pp. 2410--2414, Sept. 2022.

\end{thebibliography}

\end{document}